\documentclass[12pt,fleqn]{article}
\usepackage[DIV12]{typearea}
\usepackage{amsmath,amsfonts,amssymb,amscd}
\usepackage{graphicx}
\usepackage{cite}
\usepackage{mathrsfs}
\usepackage{bbm}
\usepackage{units}
\usepackage[small,loose]{subfigure}  
\usepackage{xcolor}
\usepackage{stmaryrd}

\usepackage{hyperref}

\newcommand{\Eqref}[1]{equation~\eqref{#1}}

\newcommand{\Tabref}[1]{table~\ref{#1}}
\newcommand{\Secref}[1]{section~\ref{#1}}
\newcommand{\Appref}[1]{appendix~\ref{#1}}

\newcommand{\SemiDirect}[0]{\ensuremath{\rtimes}}

\newcommand{\eVdist}{\kern-0.06em}

\DeclareMathOperator{\tr}{tr}

\DeclareMathOperator{\adj}{adj}

\newcommand{\D}{\mathrm{d}}
\newcommand{\I}{\mathrm{i}}
\newcommand{\e}{\mathrm{e}}

\newcommand{\SO}[1]{\ensuremath{\mathrm{SO}(#1)}}
\newcommand{\SU}[1]{\ensuremath{\mathrm{SU}(#1)}}
\newcommand{\U}[1]{\ensuremath{\mathrm{U}(#1)}}
\newcommand{\Z}[1]{\ensuremath{\mathbbm{Z}_{#1}}} 

\newcommand{\hu}{\ensuremath{H_u}}
\newcommand{\hd}{\ensuremath{H_d}}

\newcommand{\rep}[1]{\ensuremath{\boldsymbol{#1}}}
\newcommand{\crep}[1]{\ensuremath{\overline{\boldsymbol{#1}}}}
\renewcommand{\bar}[1]{\overline{#1}}

\allowdisplaybreaks[1]

\numberwithin{equation}{section}
\numberwithin{table}{section}

\def\mytitle{Non--Abelian discrete $\boldsymbol{R}$ symmetries}
\title{\mytitle}

\begin{document}

\begin{titlepage}

\begin{flushright}
 UCI-TR-2013-13\\
 TUM-HEP 895/12\\
 FLAVOR-EU 46\\
 CETUP* 12-021
\end{flushright}

\vspace*{1.0cm}

\renewcommand*{\thefootnote}{\fnsymbol{footnote}}
\begin{center}
{\Large\textbf{\mytitle}}
\renewcommand*{\thefootnote}{\arabic{footnote}}

\vspace{1cm}

\textbf{Mu--Chun Chen\footnote[1]{Email: \texttt{muchunc@uci.edu}}{}}
\\[3mm]
\emph{\small
Department of Physics and Astronomy, University of California,\\
~~Irvine, California 92697--4575, USA
}
\\[5mm]
\textbf{
Michael Ratz\footnote[2]{Email: \texttt{michael.ratz@tum.de}}{}
and 
Andreas Trautner\footnote[3]{Email: \texttt{andreas.trautner@tum.de}}{} 
}
\\[3mm]
\emph{\small
Physik Department T30, Technische Universit\"at M\"unchen, \\
~~James--Franck--Stra\ss e, 85748 Garching, Germany
}
\end{center}

\vspace{1cm}

\begin{abstract}
We discuss non--Abelian discrete $R$ symmetries which might have some
conceivable relevance for model building. The focus is on settings with
$\mathcal{N}=1$ supersymmetry, where the superspace coordinate transforms in a
one--dimensional representation of the non--Abelian discrete symmetry group. We
derive anomaly constraints for such symmetries and find that novel patterns of
Green--Schwarz anomaly cancellation emerge. In addition we show that perfect groups,
also in the non--$R$ case, are always anomaly--free. An important property of models
with non--Abelian discrete $R$ symmetries is that superpartners come in
different representations of the group. We present an example
model, based on a $\Z3\SemiDirect\Z8^R$ symmetry, to discuss generic features of
models which unify discrete $R$ symmetries, entailing solutions to the
$\mu$ and proton decay problems of the MSSM, with non--Abelian discrete flavor symmetries. 
\end{abstract}

\end{titlepage}

\section{Introduction and outline}

Despite the lack of experimental evidence for superpartners at the LHC,
supersymmetry is still one of the leading candidates for physics beyond the
standard model. The so--called $R$ symmetries, under which the superspace
coordinate $\theta$ transforms non--trivially, play an important role both in
the more formal aspects of supersymmetry as well as in model building. In the
context of $\mathcal{N}=1$ supersymmetry, the focus of the literature so
far has been on Abelian symmetries, i.e.\ either a continuous $\U1_R$ or a
discrete $\Z{N}^R$ subgroup thereof. 

It is, however, also possible to embed the $R$ symmetry in a non--Abelian
discrete symmetry group $D$. Since there is only one superspace coordinate in
the $\mathcal{N}=1$ case, $\theta$ has to furnish a one--dimensional
representation of $D$. This means that the action of $D$ on $\theta$ is Abelian,
i.e.\ a $\Z{N}^R$ symmetry. On the other hand, this $\Z{N}^R$ symmetry can
be part of a larger, in general non--Abelian symmetry group $D$.

The purpose of this study is to explore theoretical and phenomenological
properties of such symmetries. This includes anomaly constraints and possible
applications in flavor model building.

The outline of this paper is as follows. In \Secref{sec:Anomalies} we discuss
anomaly constraints and anomaly cancellation by the Green--Schwarz (GS) mechanism.
The proof that perfect groups are anomaly--free can be found in \Secref{sec:GSforNAR}.
Next, in \Secref{sec:examples} we survey possible symmetries
and discuss specifically an extension of the minimal supersymmetric standard
model (MSSM) based on a $\Z3\SemiDirect\Z8^R$ symmetry. Finally,
\Secref{sec:Summary} contains our conclusions.

\section{Anomaly constraints}
\label{sec:Anomalies}

Anomaly constraints for discrete symmetries have been analyzed using various
methods \cite{Krauss:1988zc,Ibanez:1991hv,Ibanez:1991pr,Banks:1991xj}. We will
base our discussion on the path integral approach
\cite{Fujikawa:1979ay,Fujikawa:1980eg}, which can also be applied to discrete
symmetries \cite{Araki:2006mw,Araki:2008ek}. A given symmetry operation is said
to be anomalous if it implies a non--trivial transformation of the path integral
measure. We start by reviewing the anomaly coefficients for Abelian discrete
($R$ and non--$R$) symmetries.

\subsection{Anomaly coefficients for discrete Abelian
\texorpdfstring{$\boldsymbol{R}$}{R} and
non--\texorpdfstring{$\boldsymbol{R}$}{R} symmetries}

The anomaly conditions for discrete $R$ symmetries depend on the charge of the superspace coordinate, $q_\theta$; in the
case of a non--$R$ symmetry $q_\theta=0$.  Consider an operation $\mathsf{u}$ of
order $M$, which generates a $\Z{M}$ or a  $\Z{M}^{R}$ symmetry and might or
might not be embedded in a non--Abelian symmetry group.

The superpotential transforms as
\begin{equation}
 \mathscr{W}~\to~\mathrm{e}^{2\pi\,\I\,q_\mathscr{W}/M}\,\mathscr{W}
\end{equation}
with $q_\mathscr{W}=2q_\theta$ (such that $\int\!\D^2\theta\,\mathscr{W}$ is
invariant). 
Superfields
$\Phi^{(f)}=\phi^{(f)}+\sqrt{2}\,\theta\psi^{(f)}+\theta\theta\,F^{(f)}$
transform as
\begin{equation}
 \Phi^{(f)}~\to~\mathrm{e}^{2\pi\,\I\,q^{(f)}/M}\,\Phi^{(f)}\;.
 \label{eq:abelianPhi}
\end{equation}
As a consequence, the (chiral) fermions acquire a phase
\begin{equation}
 \psi^{(f)}~\to~\mathrm{e}^{2\pi\,\I\,(q^{(f)}-q_\theta)/M}\,\psi^{(f)}\;,
 \label{eq:abelianPsi}
\end{equation}
which induces a non--trivial transformation $\mathcal D\psi^{(f)}\,\mathcal
D\bar\psi^{(f)}~\to~J^{-2}\,\mathcal D\psi^{(f)}\,\mathcal D\bar\psi^{(f)}$ of
the path integral measure with non--vanishing Jacobian. In a setting with a
non--Abelian gauge symmetry $G$ the Jacobian is given by
\begin{equation}\label{eq:Jacobian}
 J^{-2}
 ~=~
 \exp\left\lbrace \I\,\frac{2\pi}{M}\,A_{G-G-\Z{M}^R}\,
 \int\!\D^4x\,\frac{1}{32\pi^2}\,F^{b,\mu\nu} \widetilde{F}^b_{\mu\nu}\right\rbrace\;,
\end{equation}
where $F$ and $\widetilde{F}$ denote the field strength and its dual. For Abelian gauge factors and gravity one obtains analogous expressions.
In the case of a non--Abelian gauge symmetry, the mixed anomaly coefficient reads \cite{Chen:2012jg} (see also
\cite[Appendix~B]{Lee:2011dya})
\begin{equation}
 A_{G-G-\Z{M}^R}
 ~=~
 \sum_f \ell(\boldsymbol{r}^{(f)})\cdot(q^{(f)}-q_\theta)
 +q_\theta\,\ell(\adj G)
 \;.\label{eq:A_G-G-ZNR}
\end{equation}
Here, $\boldsymbol{r}^{(f)}$ denotes a representation of the gauge group $G$,
$\ell(\boldsymbol{r}^{(f)})$ is the Dynkin index of the gauge
group representation $\boldsymbol{r}$, defined as
\begin{equation}
 \delta_{ab}\,\ell(\boldsymbol{r})
 ~=~
 \tr\left[\mathsf{t}_a(\boldsymbol{r})\,\mathsf{t}_b(\boldsymbol{r})\right]\;,
\end{equation} 
and the sum goes over all fermions which transform non--trivially both under $G$
and $\mathsf{u}$. We work in conventions where $\ell(\boldsymbol{N})=\nicefrac{1}{2}$ for \SU{N}.
In this convention the Dynkin index of the adjoint is given as $\ell(\adj)=N$
for \SU{N}. In \Eqref{eq:A_G-G-ZNR}, $\ell(\adj G)=c_2(G)$ represents the contribution from the gauginos.
Here we have already allowed for $R$ symmetries, i.e.\ we include the
possibility that the superspace coordinate $\theta$ transforms non--trivially
under the operation $\mathsf{u}$. In what follows, we will mainly discuss the case of a setting with a
non--Abelian gauge symmetry $G$, but the generalization to \U1 factors and gravity is
straightforward.

Irrespective of the nature of the gauge group, all the anomaly coefficients are
only defined modulo $M/2$. Notice that for odd $M$ one can make all
odd charges even by shifting them by $M$ (cf.\ \cite{Araki:2008ek}). For such
charges, the anomaly coefficients are then only defined modulo $M$.

If a symmetry appears anomalous, this is not necessarily a sign of
inconsistency since there is the possibility of (discrete) Green--Schwarz
anomaly cancellation, which, as we will discuss in what follows, can
be employed for Abelian as well as non--Abelian discrete symmetries.

\subsection{Discrete Green--Schwarz anomaly cancellation}
\label{sec:GS}

The Green--Schwarz mechanism also works for discrete symmetries
\cite{Lee:2011dya,Chen:2012jg}. The crucial ingredient is, as usual, the
coupling of an `axion' $a$ to the field strength of the continuous gauge
symmetry
\begin{equation}\label{eq:Laxion1}
 \mathscr{L}_{\mathrm{axion}}
 ~\supset~
 -\frac{a}{8}\, F^{b} \widetilde{F}^b\;,
\end{equation}
and analogous terms for gravity (see e.g.\ \cite{Lee:2011dya} for details).
Under a discrete transformation $\mathsf{u}$  the axion undergoes a shift
\begin{equation}
 a~\to~a+\Delta^{(\mathsf{u})}\;,
\end{equation}
such that the change of $\mathscr{L}_{\mathrm{axion}}$ compensates the phase
induced by the non--trivial transformation of the path integral measure
\eqref{eq:Jacobian}. This leads to a relation between $\Delta^{(\mathsf{u})}$
and the anomaly coefficients,
\begin{equation}\label{eq:DeltaGS}
 A_{\mathsf{u}}~\equiv~ A_{G-G-\Z{M}}
 ~ = ~
 2\,\pi\,M\,\Delta^{(\mathsf{u})}
 \mod \frac{M}{2}
 \;.
\end{equation}

In principle, one can have more than one axion, in which case
\begin{equation}\label{eq:Laxion2}
 \mathscr{L}_{\mathrm{axion}}
 ~\supset~
 -\sum\limits_\alpha\frac{c_\alpha}{8}\,a_\alpha\, F^{b} \widetilde{F}^b 
\end{equation}
with some (real) coefficients $c_\alpha$. In the case of a $\Z{M}^{(R)}$
symmetry, however, there is always a unique linear combination of axions that
shifts, i.e.\ one can `diagonalize' the action on the axion fields, such that we
are back at the one--axion case. 

One can also have more than one gauge factor, i.e.\ $G=\prod_i G^{(i)}$. Then
\eqref{eq:Laxion1} generalizes to
\begin{equation}\label{eq:Laxion3}
 \mathscr{L}_{\mathrm{axion}}
 ~\supset~
 - \sum_i c_i\frac{a}{8}\, F^{(i)}_{b} \widetilde{F}^{(i)}_b 
\;.
\end{equation}
In general, the coefficients $c_i$ can be arbitrary (cf.\ 
\cite{Ludeling:2012cu}). However, in supersymmetric theories the axions are
always accompanied by a superpartner `saxion' field. In particular, in the MSSM
non--universal $c_i$ coefficients for the SM gauge factors will spoil the
beautiful picture of gauge coupling unification (see the discussion in
\cite{Chen:2012pi}). This can be avoided by demanding `anomaly universality',
which amounts to requiring
\begin{equation}
A_{G^{(i)}-G^{(i)}-\Z{M}^R}~=~\rho \mod \frac{M}{2}
~~~\forall~G^{(i)}\;,
\label{universality}
\end{equation}
and guarantees that we can use the Green--Schwarz mechanism to cancel possible
anomalies. Let us now discuss anomaly constraints on non--Abelian discrete $R$
symmetries.

\subsection{Anomaly coefficients for non--Abelian discrete \texorpdfstring{$\boldsymbol{R}$}{R} and non--\texorpdfstring{$\boldsymbol{R}$}{R} symmetries}
\label{sec:NAanomalies}

As pointed out in \cite{Araki:2006mw,Araki:2008ek}, for non--Abelian discrete
symmetries possible anomalies reside only in the Abelian parts, i.e.\ they can
be attributed to a specific generator.  Let us now focus to finite groups $D$.
Then, for each group element $\mathsf{u}\in D$ there exists an integer
$M_{\mathsf{u}}$ such that
\begin{equation}\label{eq:aNa=1}
 \mathsf{u}^{M_{\mathsf{u}}}~=~\mathbbm{1}\;,
\end{equation}
i.e.\ $\mathsf{u}$ generates a $\Z{M_\mathsf{u}}$ symmetry.   In order to verify
anomaly--freedom one has, therefore, only to check that the generators of the
group generate anomaly--free $\Z{M}$ groups.

To make this explicit, let $U_{\mathsf{u}}(\rep{d})$ be a matrix representation
of an abstract group element $\mathsf{u}\in D$ in the representation
$\rep{d}\,$. As a consequence of \eqref{eq:aNa=1},  one can always find a number
$M_\mathsf{u}$ with $U_{\mathsf{u}}(\rep{d})^{M_\mathsf{u}}=\mathbbm{1}$.
This allows us to write
\begin{equation}
 U_{\mathsf{u}}(\rep{d})
 ~=~
 \mathrm{e}^{2\pi\,\I\,\lambda_{\mathsf{u}}(\rep{d})\,/\,M_\mathsf{u}}\;,
\end{equation}
where $\lambda_{\mathsf{u}}(\rep{d})$ in general has integer eigenvalues. A
fermion charged under $D$ and transforming in a representation $\rep d^{(f)}$,
thus transforms under $\mathsf{u}$ as
\begin{equation}
 \psi^{(f)}~\to~U_{\mathsf{u}}(\rep{d}^{(f)})\,\psi^{(f)}~=~\mathrm{e}^{2\pi\,\I\,\lambda_{\mathsf{u}}(\rep d^{(f)})\,/\,M_\mathsf{u}}\,\psi^{(f)}\;.
\end{equation}
Whenever the meaning is clear from the context, we will suppress the subscript
$\mathsf{u}$ and the representation $\rep d$ for brevity. 

In the anomaly coefficient, now
\begin{equation}
 \delta^{(f)}_{\mathsf{u}}
 ~:=~
 \tr[\lambda_{\mathsf{u}}(\rep{d}^{(f)})]
 ~=~
 \frac{M_\mathsf{u}}{2\pi\,\I}\,\ln\,\det\,U_{\mathsf{u}}(\rep{d}^{(f)})\;,
\label{delta}
\end{equation}
takes the role of the discrete $\Z{M_\mathsf{u}}$ charge. This includes the
usual modulo $M$ behavior, as becomes explicit through the multi--valued
logarithm in \eqref{delta}. Nevertheless, $\delta^{(f)}$ is, in general,
\emph{not} a one--to--one replacement for an Abelian charge.  To see this
consider, for example, the relation between the transformation behavior of a
superfield $\Phi$ and the corresponding fermion, which, in analogy to equations
\eqref{eq:abelianPhi} and \eqref{eq:abelianPsi}, is given by
\begin{equation}\label{eq:fermionicrep}
\rep{d}^{(\Phi)}~=~\rep{d}^{(\theta)}\otimes\rep{d}^{(\psi)}\;.
\end{equation}
Here $\rep{d}^{(\theta)}$ denotes the representation of the superspace
coordinate $\theta$. In the case of $\mathcal{N}=1$ SUSY, $\theta$ can only
transform in a one--dimensional representation, i.e.\ $\dim(\rep
d^{(\theta)})=1$ and $\dim(\rep d^{(\psi)})=\dim(\rep d^{(\Phi)})$. Therefore,
we can express the charge of a fermion component field in terms of the
corresponding superfield charge as 
\begin{equation}
\delta^{(\psi)}~=~\delta^{(\Phi)}-\dim(\rep d^{(\Phi)})\,\delta^{(\theta)}\;.
\label{eq:fermion_charge}
\end{equation}
This illustrates that $\delta^{(f)}$ is \emph{not} a one--to--one replacement
for an Abelian charge: \Eqref{eq:fermion_charge} only reduces to the usual
addition of charges $q^R_\Phi=q^R_\theta+q^R_\psi$ for  one--dimensional
representations $\rep{d}^{(\Phi)}$.

For manifestly supersymmetric theories, it is convenient to make use of
\Eqref{eq:fermion_charge} to express the anomaly coefficients in terms of the
charges of the superfield $\delta^{(s)}$ instead of the (fermion) component
field charges $\delta^{(f)}$. 

Using this convention, let us now present the anomaly coefficients. Assume that
we have chiral superfields $\Phi^{(s)}$ which transform in representations $\rep
d^{(s)}$ of a non--Abelian discrete $R$ symmetry $D$, with charges $Q^{(s)}$
under the Abelian factors of a \U1 symmetry and as $\rep r^{(s)}$ under some
non--Abelian gauge symmetry $G$. Then, the anomaly coefficients of the Abelian,
$\mathsf{u}$--generated subgroup $\Z{M}$ of $D$ are given by
\begin{subequations}\label{eq:AnomalyCoefficients}
\begin{eqnarray}
 A_{G-G-\Z{M(\mathsf{u})}^R}
 & = &
 \sum_s \ell(\rep{r}^{(s)})\cdot\left[\delta^{(s)}-\text{dim}(\rep d^{(s)})\,\delta^{(\theta)}\right]+\ell({\adj}\,G)\cdot\delta^{(\theta)}
 \;,\label{eq:A_G-G-ZMR} \label{eq:AGGZN_na} \\
 A_{\U1-\U1-\Z{M(\mathsf{u})}^R}
 & = &
 \sum_s \left(Q^{(s)}\right)^2\,\dim(\boldsymbol{r}^{(s)})\cdot\left[\delta^{(s)}-\text{dim}(\rep d^{(s)})\,\delta^{(\theta)}\right]
 \;,\label{eq:A_U1-U1-ZMR}\\
 A_{\mathrm{grav}-\mathrm{grav}-\Z{M(\mathsf{u})}^R}
 & = &
 -21\,\delta^{(\theta)}+\delta^{(\theta)}\,\sum_G \dim(\adj G) \notag\\
 &&{} +\sum_s \dim(\boldsymbol{r}^{(s)})\cdot\left[\delta^{(s)}-\text{dim}(\rep d^{(s)})\,\delta^{(\theta)}\right]\;,
 \label{eq:A_grav-grav-ZMR}
\end{eqnarray}
\end{subequations}
where the sum goes over all chiral superfields. In the $R$ symmetry case we have
contributions not only from the matter fermions and higgsinos but also due to
possible gauge singlets, gauginos and the gravitino. The charge of the latter
two coincides with the charge of the superspace coordinate $\theta$. The anomaly
coefficients are in agreement with previous results: setting
$\delta^{(\theta)}=0$ one arrives at the coefficients for  non--$R$,
non--Abelian discrete symmetries~\cite{Araki:2008ek}, and setting 
$\delta^{(\phi)}=q^R_\phi$ and $\dim(\rep d^{(s)})=1$ leads to the coefficients
for Abelian $R$ symmetries \eqref{eq:A_G-G-ZNR} \cite{Lee:2011dya,Lee:2010gv}.

In principle, one now would have to calculate the anomaly coefficients
\eqref{eq:AGGZN_na}--\eqref{eq:A_grav-grav-ZMR} for every single group element
$\mathsf{u}\in D$ and check if they fulfill \eqref{universality}. As has been
argued in \cite{Araki:2008ek}, in the case $\rho=0$ it is enough to check
\eqref{universality} only for the generators of $D$, since if $\rho=0$ holds for
two elements $\mathsf{u},\mathsf{v} \in D$ it also holds for
$\mathsf{w}=\mathsf{u}\cdot\mathsf{v}$. This is due to the nice properties of
the determinant and the logarithm in \eqref{delta}. One has to be more careful
in the general case $\rho\neq0$ however, as will be shown in the following.

Let us assume that we have calculated the anomaly coefficients for any two group
elements $\mathsf{u}$ of order $M$ and $\mathsf{v}$ of order $N$ as
\begin{subequations}\label{eq:anomaly_coefficients}
\begin{eqnarray}
A_{\mathsf{u}} & = & \rho \mod \frac{M}{2}
\;, \\
A_{\mathsf{v}} & = & \sigma \mod \frac{N}{2}
\;.
\end{eqnarray}
\end{subequations}
The anomaly coefficient of a third group element $\mathsf{w}=\mathsf{u}\cdot
\mathsf{v}$ of order $L$ then is given by\footnote{In the most general case
(where we do not assume anything about the permuting properties or relative
orders of $\mathsf{u}$ and $\mathsf{v}$) we cannot say much about the relation
of $L$, $M$ and $N$.}
\begin{eqnarray}
 A_{\mathsf{w}}
& = &\sum_f \ell(\rep{r}^{(f)})\,\delta^{(f)}_{\mathsf{w}} +
 \ell({\adj}\,G)\,\delta^{(\theta)}_{\mathsf{w}} \nonumber\\
& = &\sum_f
 \ell(\rep{r}^{(f)})\left(\frac{L}{M}\,\delta^{(f)}_{\mathsf{u}} +
 \frac{L}{N}\,\delta^{(f)}_{\mathsf{v}}\right)+\ell({\adj}\,G)\left(
 \frac{L}{M}\,\delta^{(\theta)}_{\mathsf{u}}+\frac{L}{N}\,\delta^{(\theta)}_{\mathsf{v}}\right)
\nonumber \\
& =&  
	  \frac{L}{M}\left(\rho \mod \frac{M}{2}
	  \right)
	  +\frac{L}{N}\left(\sigma \mod \frac{N}{2}
	  \right)\;.\label{eq:combinedanomaly}
\end{eqnarray}
We can now distinguish three possible cases:
\begin{enumerate}
\item Neither $\mathsf{u}$ nor $\mathsf{v}$ generates an anomalous symmetry, i.e.\ $\rho=\sigma=0$. 
We recover the trivial case as treated in \cite{Araki:2008ek}. The symmetry generated by $\{\mathsf{u},\mathsf{v}\}$ is anomaly--free.
\item Without loss of generality, only $\mathsf{u}$ generates an anomalous symmetry, i.e.\ $\rho\neq0=\sigma$. 
It follows that also $\mathsf{w}=\mathsf{u}\cdot \mathsf{v}$ is anomalous, with an anomaly coefficient
\begin{equation}
A_{\mathsf{w}}~=~L\,\left(\frac{\rho}{M}\mod \frac{1}{2}
\right)\;.
\end{equation}
\item Both $\mathsf{u}$ and $\mathsf{v}$ generate anomalous symmetries. The anomaly
coefficient for $\mathsf{w}$ is
\begin{equation}
A_{\mathsf{w}}~=~L\left[
\left(\frac{\rho}{M} + \frac{\sigma}{N}\right) \mod\frac{1}{2}\right]\;.
\label{an_combined}
\end{equation}
In this case, even though $\mathsf{u}$ and $\mathsf{v}$ appear anomalous,
$\mathsf{w}$ might not. Note also the special case $\mathsf{u}=\mathsf{v}$ where
$\mathsf{w}=\mathsf{u}^2$ appears anomalous if and only if
$\nicefrac{4\rho}{M}\notin\mathbbm{Z}$.  
\end{enumerate} 
A generalization of this
discussion to three or more generators is possible in a straightforward way.

\subsection{Green--Schwarz mechanism for non--Abelian discrete symmetries}
\label{sec:GSforNAR}

In principle, the cancellation mechanism for Abelian discrete symmetries also
works for the Abelian subgroups of non--Abelian symmetries. There are, however,
some additional relations constraining possible axion transformations under the
symmetry group. Consider two operations, $\mathsf{u}$ and $\mathsf{v}$, in $D$.
In general, those will induce shifts 
\begin{subequations}
\begin{eqnarray}
 \mathsf{u}~:~ a & \to & a+\Delta^{(\mathsf{u})}\;,\\
 \mathsf{v}~:~ a & \to & a+\Delta^{(\mathsf{v})}\;.
\end{eqnarray}
\end{subequations}
In particular, the action of these shifts on the axion is Abelian --- in other
words: the chiral superfield containing the axion as complex phase can only
transform in a one--dimensional representation of our symmetry. As a
consequence, axions are not allowed to shift under so--called commutator
elements of the symmetry group\footnote{Here the group--theoretical definition
of the commutator (cf.\  \cite{Ramond:2010zz})
$[\mathsf{u},\mathsf{v}]:=\mathsf{u}\,\mathsf{v}\,\mathsf{u}^{-1}\,\mathsf{v}^{-1}$
is used.} $\mathsf{x}:=[\mathsf{u},\mathsf{v}]$, since for such elements all
fermion charges 
\begin{equation}
 \delta_\mathsf{x}~=~\frac{M}{2\pi\,\I}{\,\ln\,\det\,}U_\mathsf{x}\;,
\label{delta_com}
\end{equation}
and therefore the anomaly coefficients
\eqref{eq:AGGZN_na}--\eqref{eq:A_grav-grav-ZMR} trivially vanish. This
immediately follows from the definition of a commutator element, whose
representations always can be written as
\begin{equation}
 U_\mathsf{x}~=~U_{\mathsf{u\,v\,u^{-1}\,v^{-1}}}~=~U_{\mathsf{u}}\,U_{\mathsf{v}}\,U_{\mathsf{u}^{-1}}\,U_{\mathsf{v}^{-1}}~=~U_{\mathsf{u}}\,U_{\mathsf{v}}\,{U_{\mathsf{u}}}^{-1}\,{U_{\mathsf{v}}}^{-1}\;,
\end{equation}
which leads to a vanishing charge \eqref{delta_com} and thus to vanishing anomaly
coefficients. By noting that only the one--dimensional representations
transform trivially under commutator elements, it is clear that axions can only
transform as one--dimensional representations.

We would like to remark that, for the same reason, perfect groups, which are
generated by commutator elements only, \emph{always} are anomaly--free.
Nevertheless, since they do not possess non--trivial one--dimensional
representations, perfect groups are not relevant to this work.

Let us now discuss the cancellation of anomalies for the whole non--Abelian
group. Consider two generating elements $\mathsf{u}$ and $\mathsf{v}$ of order
$M$ and $N$ with their respective anomaly coefficients (cf.\ 
\eqref{eq:anomaly_coefficients}). The combined operation
$\mathsf{u}\cdot\mathsf{v}$ is assumed to have order $L$, and the anomaly
coefficient of the combined operation is given by
\begin{equation}
A_{\mathsf{u}\cdot\mathsf{v}}~=~\omega\mod \frac{L}{2}
\;.
\end{equation}
As shown in \Eqref{eq:combinedanomaly}, the combined anomaly coefficient can be
rewritten as a non--trivial sum of the single anomaly coefficients. Let us now
check whether it is always possible to cancel the combined anomaly. To do so, we
impose an axion shift
\begin{equation}
 \mathsf{u}\cdot\mathsf{v}~:~ a~ \to ~ a+\Delta^{(\mathsf{u}\cdot\mathsf{v})}\;,
\end{equation}
which, due the Abelian nature of the axion transformation, must be given as
\begin{equation}
 \Delta^{(\mathsf{u}\cdot\mathsf{v})}~=~\Delta^{(\mathsf{u})}+\Delta^{(\mathsf{v})}\;.
\label{abelian_shift}
\end{equation}
The condition for the cancellation of the combined anomaly, in analogy to
\eqref{eq:DeltaGS}, is
\begin{equation}
A_{\mathsf{u}\cdot\mathsf{v}}~=~2\,\pi\,L\,\Delta^{(\mathsf{u}\cdot\mathsf{v})}
\mod\frac{L}{2}
\;,
\end{equation}
which can be rewritten as
\begin{eqnarray}
 A_{\mathsf{u}\cdot\mathsf{v}} 
  &\stackrel{(\ref{abelian_shift})}{=}& 2\,\pi\,L\,\left(\Delta^{(\mathsf{u})}+\Delta^{(\mathsf{v})}\right) \mod \frac{L}{2}
  \nonumber\\
  &\stackrel{(\ref{eq:DeltaGS})}{=}& \frac{L}{M}\,\left(\rho\mod \frac{M}{2}
  \right)+\frac{L}{N}\,\left(\sigma\mod \frac{N}{2}
  \right)\;.
\end{eqnarray}
But this is exactly the same as \eqref{eq:combinedanomaly}. This means we do not
have any constraints on the anomalies of combined elements, or in other words,
if the single (Abelian) anomalies of the generator elements are vanishing (with
or without employing the Green--Schwarz mechanism), the whole group is
anomaly--free.

\section{Non--Abelian discrete \texorpdfstring{$\boldsymbol{R}$}{R} symmetries in the MSSM}
\label{sec:examples}

\subsection{Symmetry search}
\label{sec:symmetry search}

In what follows, we will discuss specific examples of non--Abelian discrete $R$
symmetries in the context of the MSSM. The non--Abelian discrete $R$ symmetry
will, in general, act non--trivially on flavor space. We will assume it to be
partly broken by flavon VEVs at a high scale, thus giving rise to a specific
flavor structure. On the other hand, since we wish not to break supersymmetry at
a high scale $\Lambda$, we require the $R$ symmetry subgroup to be unbroken. 
Specifically, we will focus on settings in which there is a residual $\Z4^R$
symmetry \cite{Babu:2002tx}, which has recently been shown to be the unique
Abelian discrete $R$ symmetry which allows us to solve the $\mu$ problem and
commutes with SO(10) in the matter sector
\cite{Lee:2010gv,Lee:2011dya,Chen:2012jg}. An unbroken $\Z{2}$ subgroup of this
symmetry coincides with $R$ parity.

We further demand that, after breaking the
flavor symmetry $D$ down to the residual $\Z4^R$ symmetry, matter fields have
$R$ charge 1 and Higgs fields charge 0. This is because we assume a hierarchy
between $\Lambda$ and the scale of $\Z4^R$ breaking, which is given by the
gravitino mass $m_{\nicefrac{3}{2}}$. In this case, a family--dependent $\Z4^R$ charge
assignment implies unrealistic mixing angles. Hence, in order to be consistent with this
charge assignment while allowing for correlations in family space, the
non--Abelian discrete $R$ symmetry is required to have a multiplet
representation whose components transform equally under the $\Z4^R$ subgroup.
In particular, this requires that the center of the group contains the $\Z4^R$.
One can see this with the help of an explicit representation: consider the
representation matrix of the generating element of the $\Z4^R$ subgroup in a
basis in which it is diagonal. Since each component is required to transform
equally under the subgroup, this matrix must be proportional to the unit matrix,
therefore commuting with all other matrices of the representation, and thus a
representation of an element of the center.

To summarize, we survey non--Abelian discrete $R$ symmetries which satisfy the
following criteria:
\begin{enumerate}
 \item the symmetry contains, and can be spontaneously broken down to a $\Z4$
   symmetry by a multiplet VEV;
 \item the symmetry contains a one--dimensional representation (for
  $\theta$), which transforms non--trivially also under the unbroken subgroup;
 \item the residual $\Z4$ subgroup is part of the center of the symmetry group.
\end{enumerate}
We have conducted a symmetry search in the {\footnotesize SMALLGROUPS} library
of the {\footnotesize GAP} system for computational discrete algebra
\cite{GAP4}. The results for the groups up to order 48 are shown in table
\ref{tab:symsearch}.
\begin{table}[ht!]
\centering
\begin{tabular}{c|l|l}
 $\mathcal O(D)$ & Structure description & ID  \\
\hline  
24 & $\Z3\rtimes\Z8$ & SG(24,1) \\
24 & $\mathrm{S}_3\times\Z4$ & SG(24,5) \\
32 & $(\Z8\times\Z2)\rtimes\Z2$ & SG(32,5) \\
32 & $(\Z4\times\Z4)\rtimes\Z2$ & SG(32,11) \\
32 & $\Z8\rtimes\Z4$ & SG(32,12) \\
32 & $D/\Z4 = \mathrm{D}_{8}$ & SG(32,15) \\
32 & $(\Z4\times\Z4)\rtimes\Z2$ & SG(32,24) \\
40 & $\Z5\rtimes\Z8$ & SG(40,1) \\
40 & $\Z4\times\mathrm{D}_{10}$ & SG(40,3) \\
48 & $\Z{24}\rtimes\Z2$ & SG(48,5) \\
48 & $(\Z3\rtimes\Z8)\times\Z2$ & SG(48,6) \\
48 & $(\Z3\rtimes\Z8)\rtimes\Z2$ & SG(48,7) \\
48 & $(\Z3\rtimes\Z4)\times\Z4$ & SG(48,8) \\
\vdots & \vdots & \vdots\\
\end{tabular}
\caption{Result of the {\footnotesize GAP} scan, showing groups consistent with
the requirements stated in the text up to order 48. We give order, name and/or
structure description of the group as well as the {\footnotesize SMALLGROUPS}
library ID of {\footnotesize GAP}.}
\label{tab:symsearch}
\end{table}
The smallest groups which fulfill all requirements are $\Z3\SemiDirect\Z8$ and
${\mathrm S}_3\times\Z4$.  The latter group contains the well known ${\mathrm
S}_3$, on which several working GUT flavor models are based 
\cite{Chen:2004rr,Morisi:2005fy,Dermisek:2005ij,Caravaglios:2005gf,Teshima:2005bk,Haba:2005ds,Picariello:2006sp,Mohapatra:2006un,Mohapatra:2006pu,Feruglio:2007hi}. 
Nevertheless, regarding an $R$ symmetric extension of the MSSM, it would just be
the trivial extension of any of the known ${\mathrm S}_3$ models by a $\Z4^R$. 
Such models should not concern us here. We will focus our considerations on the
other possible lowest order group, namely $\Z3\SemiDirect\Z8$.  In contrast to
${\mathrm S}_3$, we are not aware of any existing flavor model based on this
group.  For this reason, we have stated the necessary group theoretical details
in \Appref{appsec:group}.

\subsection{\texorpdfstring{$\boldsymbol{\Z3\SemiDirect\Z8^R}$}{Z3xZ8} extension of the MSSM}
\label{sec:model}

Let us now discuss an example model for non--Abelian discrete $R$
symmetries, which also act in flavor space, based on the particle spectrum of
the the MSSM. Taking grand unification seriously, we will arrange the matter
fields in \SU5 multiplets. We further impose the condition of `anomaly
universality' (cf.\ the discussion in \cite{Chen:2012pi}) such that discrete
anomalies can be cancelled by the GS mechanism without spoiling gauge coupling
unification.

We will focus our discussion on the generic features of non--Abelian discrete
$R$ symmetry extensions rather than trying to enforce an entirely correct
phenomenology. Thus, in the spirit of minimalism, we spare additional Abelian
discrete `shaping' symmetries and flavons other than the ones which are
essential to symmetry breaking. The explicit construction of a possibly fully
realistic model is left for future work. In the present work, we will employ a
minimal example model to discuss the consistent assignment of representations, 
generalities of the symmetry breaking and VEV alignment, the construction of
Yukawa coupling and mass matrices, and the explicit calculation of anomaly
coefficients.

\paragraph{$\boldsymbol{\Z3\rtimes\Z8}$ and the $\boldsymbol{\Z4}$ subgroup.}
The group $\Z3\rtimes\Z8$ is generated by the two elements $\mathsf{u}$
and $\mathsf{v}$, which fulfill
\begin{equation}\label{eq:presentation}
 \Z3\SemiDirect\Z8
 ~=~
 \left\langle \mathsf{u},\mathsf{v}\;; ~ 
 \mathsf{u}^3=\mathsf{v}^8=\mathbbm{1}\;,
 ~\mathsf{v}\,\mathsf{u}\,\mathsf{v}^{-1}
 =\mathsf{u}^{-1} \right\rangle\;.
\end{equation}
The group is of order 24, it has 8 one--dimensional representations that we
label as  $\rep1_i~(i=1\dots8)$ and 4 doublets, denoted by
$\rep2_j~(j=1\dots4)$. A more detailed discussion of the group is deferred to
\Appref{appsec:group}. As, by assumption, a $\Z4$ subgroup of $\Z3\rtimes\Z8$
will survive down to the SUSY breaking scale, we list the behavior of
irreducible representations under this $\Z4$ subgroup (which is the one
generated by $\mathsf{v}^2$) in \Tabref{tab:decompo}.
\begin{table}[t]
\[
\begin{split}
&\Z3\SemiDirect\Z8&~~  &\rep{1}_2& &\rep{1}_3& &\rep{1}_4& &\rep{1}_5& &\rep{1}_6& &\rep{1}_7& &\rep{1}_8& &\rep{2}_1& &\rep{2}_2& &\rep{2}_3& &\rep{2}_4&\\
 &&~~ &\downarrow& &\downarrow& &\downarrow& &\downarrow& &\downarrow& &\downarrow& &\downarrow& &\downarrow& &\downarrow& &\downarrow& &\downarrow&\\
&\Z4&~~ &\rep{1}& &\rep{1}^{\prime\prime}& &\rep{1}^{\prime\prime}& &\rep{1}^{\prime}& &\rep{1}^{\prime\prime\prime}& &\rep{1}^{\prime\prime\prime}& &\rep{1}^{\prime}& \rep{1}^{\prime\prime}&\oplus\rep{1}^{\prime\prime}& \rep{1}&\oplus\rep{1}& \rep{1}^{\prime\prime\prime}&\oplus\rep{1}^{\prime\prime\prime}& \rep{1}^{\prime}&\oplus\rep{1}^{\prime}&
\end{split}
\]
\caption{Branching rules for $\Z3\SemiDirect\Z8\to\Z4$.}
\label{tab:decompo}
\end{table}
Here $\rep{1}^{\prime},\rep{1}^{\prime\prime}$ and
$\rep{1}^{\prime\prime\prime}$ label the representations of $\Z4$, with the
number of primes specifying the corresponding charge. For example, matter fields
and the superspace coordinate $\theta$ will transform in the $\rep{1}^{\prime}$
representation. As $\theta$ transforms
non--trivially, the residual symmetry is an (order four) $R$ symmetry, denoted
as $\Z4^R$ in what follows. We observe that the $\rep{2}_2$ contains twice the
trivial singlet of $\Z4^R$. Thus a $\rep{2}_2$ VEV in \emph{any} direction can
break $\Z3\SemiDirect\Z8^R\rightarrow\Z4^R$, as desired. Note also
that a $\rep2_2$ VEV aligned in the $(1,0)$ direction (in the basis specified in
\eqref{eq:matrix_rep_Z8}) would break $\Z3\SemiDirect\Z8^R\rightarrow\Z8^R$.

\paragraph{Charge assignment.}
From the requirement that $\theta$ carries $\Z4^R$ charge\footnote{As discussed
for instance in \cite{Chen:2012jg}, any $\Z{M}^R$ symmetry solution to the $\mu$
problem requires $M=4\times\mathbbm{N}$ and $q_\theta=\nicefrac{M}{4}$.} 1 and
the breaking pattern of $\Z3\SemiDirect\Z8^R\rightarrow\Z4^R$ in
\Tabref{tab:decompo}, we infer that $\theta$ has to transform as a
$\rep{1}_5$ (or as a $\rep{1}_8$, which would make no difference), the
Higgs fields as $\rep{1}_1$ or $\rep{1}_2$, and matter can be assigned to
$\rep1_5$, $\rep1_8$ or $\rep2_4$ under $\Z3\SemiDirect\Z8^R$. For a
general $\Z{M}^R$ symmetry, in order to be anomaly universal,
\Eqref{eq:A_G-G-ZNR} applied to the non--Abelian gauge groups immediately leads
to the requirement
\begin{equation}
q_{\hu}+q_{\hd}~=~4\,q_\theta\mod M\;,
\label{eq:higgs_condition}
\end{equation}
for the Higgs charges (cf.\ e.g.\ \cite{Lee:2010gv}).  Applying this to the
$\Z8^R$ subgroup, we conclude that the Higgs fields have to transform in
different representations.  This will be important also in the explicit
computation of anomaly coefficients in \Secref{sec:model_anomalies}.

To accomplish the breaking of the family symmetry
$\Z3\SemiDirect\Z8^R\rightarrow\Z4^R$, we need at least one additional, SM
singlet degree of freedom which transforms as a $\rep{1}_2$ or $\rep2_2$ and
acquires a VEV. We therefore introduce two of such
`flavons', $\phi$ and $\chi$, transforming as $\rep{1}_2$ and $\rep2_2$,
respectively.

Different assignments either lead to a different breaking of
$\Z3\SemiDirect\Z8^R$ or to unfeasible $\Z4^R$ charge assignments.  The
assignment we choose in accordance with all imposed requirements is listed in
\Tabref{tab:assignment}.  Of course, variations of the assignment of the matter
and Higgs fields are possible. We have chosen our example such that one gets a
glimpse on the variety of possible (leading order) mass matrix structures.
\begin{table}[t]
\begin{center}
\begin{tabular}{|c|c|c|c|c|c|c|c|c|}
\hline\vphantom{$\left(\begin{array}{c}1\\ 1\end{array}\right)$}
 $\left( Q, \bar{U}, \overline{E}\right)_{1,2}$ & $\left( Q, \bar{U}, \bar{E}\right)_{3}$ & $\left(\bar{D}, L \right)_{1,2} $ & $\left(\bar{D},L\right)_{3}$ & 
   $ \hu$ & $\hd$ & $\chi$ & $\phi$ & $\theta$ \\
\hline
 $\rep{1}_5$ & $\rep{1}_8$ & $\rep{2}_4$ & $\rep{1}_5$ & $\rep1_1$ & $\rep1_2$ & $\rep2_2$ & $\rep1_2$ & $\rep1_5$ \\
\hline
\end{tabular}
\end{center}
\caption{Transformation of MSSM fields under $\Z3\SemiDirect\Z8^R$. The notation
for the MSSM fields is standard, $\theta$ is the superspace coordinate, $\chi$
and $\phi$ are (MS)SM singlet flavons.}
\label{tab:assignment}
\end{table}
There is one peculiar difference here with respect to traditional flavor models:
since we are dealing with an $R$ symmetry,  the allowed superpotential terms may
not be neutral but have to be charged instead. Since $\theta$ resides in a
$\rep{1}_5$,  the charge of the superspace integral measure is
$\rep{1}^*_5\otimes\rep{1}^*_5=\rep{1}^*_4=\rep{1}_3$.  Therefore,
superpotential terms have to transform as $\rep{1}_4$. We wish to point out
that, given the non--trivial transformation of $\theta$, fermions and bosons
furnish \emph{different} representations under the flavor group. For instance,
if a superfield transforms as $\rep{2}_4$, then the scalar components also
furnishes this representations, but the fermions have to transform as
$\rep{2}_2$ if $\theta$ transforms as $\rep{1}_5$ or $\rep{1}_8$. 

\paragraph{Spontaneous breaking $\boldsymbol{\Z3\SemiDirect\Z8^R\to\Z4^R}$.}
From the branching rules (cf.\ \Tabref{tab:decompo}) and the charges of the
flavons, it is clear that a non--trivial VEV of either $\phi$ or $\chi$ will
break $\Z3\SemiDirect\Z8^R\to\Z4^R$.  In the general case, for the generation of
potentially realistic fermion masses, we need to switch on both,
$\langle \phi \rangle$ and $\langle \chi \rangle$. 
Note that, in order to achieve the breaking to the $\Z4^R$, the doublet $\chi$
does not have to be aligned in any way since both components of the doublet
transform trivially under this subgroup. The fact that multiplet VEVs do not
have to be aligned for a desirable breaking pattern is a generic feature of the
non--Abelian discrete $R$ symmetries under discussion as can be
inferred from the fact that the unbroken $\Z4^R$ is required to be in the center
of the non--Abelian group.

In the case of $\Z3\SemiDirect\Z8^R$, however, an alignment of $\langle \chi
\rangle$ along the $(1,0)$ direction can arise due to the presence of a single
additional field $\xi$ transforming as $\rep1_4$ under the $R$ symmetry and
trivially under all other symmetries. At the renormalizable level, $\xi$
couples only linearly to the flavon fields and does not possess couplings to the
MSSM fields in the superpotential $\mathscr{W}$. Therefore, $\xi$
automatically possesses the typical characteristics of a `driving' field. In
order to study the alignment, let us parameterize the VEVs as $\langle \chi
\rangle ~=~v\,(\cos{\theta_\chi},\sin{\theta_\chi})^T$ and $\langle \phi \rangle
~=~ v\,r_\phi$. Requiring SUSY to be unbroken at the flavor scale, one obtains
the $F$--term condition
\begin{equation}\label{eq:Fterm}
 0~\stackrel{!}{=}~\left.
\frac{\partial\,\mathscr{W}}{\partial\xi}\right|_{\substack{\phi\rightarrow\langle\phi\rangle
\\
\chi\rightarrow\langle\chi\rangle}}
~=~-M^2+g_1\,v^2\,\left(2\,\cos^2{\theta_\chi}-1\right)+g_2\,v^2\,{r_\phi}^2\;,
\end{equation}
where we take $M^2,g_1,g_2>0$. What is crucial for the alignment is a
choice of parameters such that there is a relative sign difference between  the
first and second terms. As one can check from \eqref{eq:Fterm}, $v$ and $r_\phi$
are minimized for $\theta_\chi=0$, i.e.\ $\langle \chi \rangle\propto(1,0)$. 
This corresponds to a breaking $\Z3\SemiDirect\Z8^R\to\Z8^R$ which would lead to
the vanishing of two mixing angles since the residual $\Z8^R$ symmetry is family
dependent. We see that \emph{this alignment has to be avoided} in order to
obtain a correct phenomenology. However, a mild suppression of the
leading--order contribution is enough to generate a small misalignment from the
next--to--leading order terms of the superpotential, resulting in
$\theta_\chi\approx\delta$, hence, modifying the VEV to $\langle \chi
\rangle\propto(1,\delta)$. This then leads to a breaking
$\Z3\SemiDirect\Z8^R\rightarrow\Z4^R$ with a slightly broken and hence
approximate $\Z8^R$.  The small misalignment could, for instance, help to
explain the small mixing to the third generation. In what follows, we will
work with the VEVs 
\begin{equation}
 \langle \chi \rangle ~=~v\, \binom{1}{\delta}
 \quad\text{and}\quad
 \langle \phi \rangle ~=~ v\,r_\phi\;.
\label{vevs2}
\end{equation}

\paragraph{Effective fermion mass matrices.}
We now use the direct product rules and the tensor structure of the
decomposition \eqref{eq:tensorsZ8B} to identify terms consistent with all
symmetries.  The effective neutrino mass operator is given by
\begin{eqnarray}
\mathscr{W}_\nu^\mathrm{eff} &=&(\hu\,L^g)\,\kappa_{gf}\,(\hu\,L^f)\nonumber \\
 & = & \frac{v_u^2}{\Lambda_\nu}\left\lbrace x_1\,\left( L_1\,L_1 - L_2\,L_2
 \right) +  x_3\,L_3\,L_3 + 2\, x_4\,\frac{L_3}{\Lambda}\left( \chi_1\,L_2 -
 \chi_2\,L_1 \right) \right. \nonumber\\
 &&{}  \left. \hphantom{\frac{v_u^2}{\Lambda_\nu}~~ } + x_2\,\left[ \frac{\chi_1}{\Lambda}\left( L_1\,L_1 + L_2\,L_2 \right) + \frac{\chi_2}{\Lambda}\left( L_1\,L_2 + L_2\,L_1 \right) \right] \right\rbrace\;,
\end{eqnarray} 
where we have introduced dimensionless coupling coefficients $x_i$ (in the
following also $y_i,z_i$), the see--saw scale $\Lambda_\nu$, as well as the
flavor scale $\Lambda$, and set the Higgs fields to their VEVs.  Terms involving
more flavons are of higher order in $\varepsilon:=\nicefrac{v}{\Lambda}$ and are
not discussed here. Setting the flavons to their VEVs, the emerging structure of
the effective neutrino mass matrix is 
\begin{equation}
\kappa~=~\frac{v_u^2}{\Lambda_\nu}
\begin{pmatrix}
 x_1+ x_2\,\varepsilon & x_2\,\varepsilon\,\delta & -x_4\,\varepsilon\,\delta \\
 x_2\,\varepsilon\,\delta & -x_1+x_2\,\varepsilon & x_4\,\varepsilon \\
 -x_4\,\varepsilon\,\delta & x_4\,\varepsilon & x_3
\end{pmatrix}\;.
\label{eq:Mnu}
\end{equation}

The effective charged lepton mass is constrained to the form
\begin{eqnarray}
 \mathscr{W}_e & =& \bar{E}^fY^{(e)}_{fg}(\hd\,L^g)\, \nonumber\\
 &= & v_d \left\lbrace y_1\,\frac{\bar{E}_1}{\Lambda}\left( \chi_1\,L_1-\chi_2\,L_2 \right)+ y_2\,\frac{\bar{E}_2}{\Lambda}\left( \chi_1\,L_1-\chi_2\,L_2 \right) + \right. 
 \nonumber\\
 & & \left. \hphantom{v_d~~} y_3\,\frac{\bar{E}_3}{\Lambda}\left( \chi_1\,L_2-\chi_2\,L_1 \right) + y_4\,\frac{\phi}{\Lambda}\,\bar{E}_1\,L_3 + y_5\,\frac{\phi}{\Lambda}\,\bar{E}_2\,L_3 + y_6\,\bar{E}_3\,L_3\right\rbrace\;,
\end{eqnarray}
resulting in the structure 
\begin{equation}
Y^{(e)}~=~v_d
\begin{pmatrix}
  y_1\,\varepsilon & -y_1\,\varepsilon\,\delta &  y_4\,\varepsilon\,r_\phi \\
  y_2\,\varepsilon & -y_2\,\varepsilon\,\delta &  y_5\,\varepsilon\,r_\phi \\
  -y_3\,\varepsilon\,\delta & y_3\,\varepsilon &  y_6\, 
\end{pmatrix}\;.
\label{eq:Ye}
\end{equation}
As usual for settings with \SU5 relations, we have $Y^{(e)}\sim
{Y^{(d)}}^T$, which immediately fixes the structure of the down--quark
Yukawa coupling.
The up--quark Yukawa coupling has less structure since only
one--dimensional representations are contracted. We find
\begin{equation}
Y^{(u)}=v_u\,
\begin{pmatrix}
  z_1 &  z_2  &  z_5\,\varepsilon\,r_\phi \\
  z_3 &  z_4  &  z_6\,\varepsilon\,r_\phi \\
  z_7\,\varepsilon\,r_\phi & z_8\,\varepsilon\,r_\phi & z_9\,
\end{pmatrix}\;.
\label{eq:Yu}
\end{equation}

As already mentioned, it is possible to have variations of the charge assignment
in \Tabref{tab:assignment} which are consistent with all imposed requirements.
Besides permutation in the family indices, such variations can only lead to mass
matrices that are similar in  structure to the ones of the example shown above.
More precisely, one could, instead of the \crep{5}--plets, combine two
generations of the \SU5 \rep{10}--plets to a doublet,  leading to a similar but
transposed structure for $Y^{(e)}$ and $Y^{(d)}$, and to a swap in the structure
of $Y^{(u)}$ and $\kappa$.  Alternatively, also a setup in which two generations
each of the \crep{5} and \rep{10}--plets get combined to doublets is possible,
which is the only possibility in case of an \SO{10} GUT. In this case, all mass
matrices will take a form similar to \eqref{eq:Mnu}.

\paragraph{Model Phenomenology.}
Even though we did not arrange our model to fit the experimental data, let us
comment on the resulting phenomenology as it would be a starting point for the
construction of possibly realistic models. Without imposing any additional
symmetries, there are unsuppressed tree--level contributions to the mass
matrices next to suppressed effective terms. 
As in other flavor models with non--Abelian discrete symmetries, it is clear
that also in this case one needs to introduce further symmetries, such as, the
so--called shaping symmetries or a \U1 of the Froggatt--Nielsen type, in order
to obtain a completely natural and realistic model with hierarchical masses. For
the particular model considered here, a Froggatt--Nielsen symmetry with
$\lambda\sim\theta_c\sim0.2$ may be used to explain the hierarchy among the
parameters
\begin{subequations}
\begin{eqnarray}
y_1~:~y_2~:~y_3~:~y_6 & = & \lambda^4~:~\lambda^2~:~\lambda^0~:~\lambda^1\;, \\ 
z_1~:~z_4~:~z_9 &= &\lambda^8~:~\lambda^4~:~\lambda^0\;,
\end{eqnarray}
\end{subequations}
which can lead to a good agreement with the data as has been checked numerically
using the {\footnotesize MPT} package \cite{Antusch:2005gp}. However, as this is
just a toy model with more parameters than observables, we refrain from fitting the model predictions to data. Yet our
discussion shows that 
viable flavor models can, in principle, arise from non--Abelian discrete $R$ symmetries,
analogous to case of non--$R$, non--Abelian discrete symmetries 
(see e.g.\
\cite{Altarelli:2010gt,Ishimori:2010au,King:2013eh} for reviews). 
This in turn affords the possibility of having a simultaneous solution to 
the $\mu$ problem and the flavor problem.
In what follows we will use
the toy model as a basis for an explicit calculation of the anomaly
coefficients.

\paragraph{Anomalies of the $\boldsymbol{\Z3\SemiDirect\Z8^R}$ Model.}
\label{sec:model_anomalies}
Finally, we can use formulae \eqref{eq:AnomalyCoefficients}
to calculate the
$R$--gauge--gauge anomaly coefficients of the $\Z3\SemiDirect\Z8^R$ model. For
this, we first have to calculate the charges of every representation. For the
symmetry treated here, there are only two generators $\mathsf{u}$ and
$\mathsf{v}$. The representation matrix $U$ equals the respective character for
the one dimensional representations, and can be read off from equations
\eqref{eq:matrix_rep_Z8_A}--\eqref{eq:matrix_rep_Z8_B} for the two dimensional
representations. Since $\text{det}~U_{\mathsf{u}}=1$ for all representations, the
symmetry generated by $\mathsf{u}$ is trivially anomaly--free and we only have to
care about $\mathsf{v}$. The $\delta$ charges \eqref{delta} for all relevant
conjugacy classes are given in table \ref{tab:charges}.
\begin{table}[t]
\centering
\begin{tabular}{c|cccccc}
 $\rep d^{(\Phi)}$ & $\rep{1}_1$ & $\rep{1}_2$ & $\rep{1}_5$ & $\rep{1}_8$ & $\rep2_2$ & $\rep2_4$ \\
\hline
$\delta^{(\Phi)}_{\mathsf{v}}$ & 0 & $4$ & $5$ & $1$ & $4$ & $6$ \\
$\delta^{(\Phi)}_{\mathsf{v^2}}$ & 0 & 0 & $1$ & $1$ & 0 & $2$ \\
\end{tabular}
\caption{Charges of fields under the $\mathsf{v}$ and $\mathsf{v}^2$ generated
subgroups of $\Z3\rtimes\Z8^R$, computed with \Eqref{delta}. The charges
are only defined modulo $M_{\mathsf{v}}=8$ and $M_{\mathsf{v^2}}=4$,
respectively.}
\label{tab:charges}
\end{table}
Here it pays off that we have expressed the anomaly coefficients in terms of the
superfield charges via \eqref{eq:fermion_charge}, such that in order to
find the charges relevant for the anomaly coefficient we do not have to work
out the representations of the fermion component fields and their respective
charges, but instead take the superfield charge from \Tabref{tab:charges}
and subtract the charge of $\theta$ times the dimensionality of the respective
superfield's representation.  We use the modulo $M$ freedom to shift all
charges to positive values as a convention. Putting everything together, we find
for the anomaly coefficients of the discussed model under the $\mathsf{v}$
generated subgroup of the discrete non--Abelian family symmetry
$\Z3\rtimes\Z8^R$ the expressions
\begin{subequations}
\begin{align}
&A_{\SU3-\SU3-\Z{8(\mathsf{v})}^R} ~=~ \frac{1}{2}\left\lbrace \left[ 2+1
\right]\cdot4+\left[1\right]\cdot4\right\rbrace+3\cdot5~=~3\mod4\;,\\ \label{su2anomaly}
&A_{\SU2-\SU2-\Z{8(\mathsf{v})}^R} ~=~ \frac{1}{2}\left\lbrace \left[ 3 \right]
\cdot4 + \left[1\right]\cdot4+ 3 + 7
\right\rbrace+2\cdot5~=~3\mod4\;,\\
\begin{split}
&A_{\U1-\U1-\Z{8(\mathsf{v})}^R} ~=~ \frac{3}{5}\left\lbrace \left[ 3\cdot2\cdot\left(\frac{1}{6}\right)^2+3\cdot\left(\frac{2}{3}\right)^2+(1)^2\right] \cdot 4~+ \right. \\
& \left.\left[ 3\cdot\left(\frac{1}{3}\right)^2  +2\cdot\left(\frac{1}{2}\right)^2\right]\cdot4+2\cdot\left(\frac{1}{2}\right)^2\cdot
3+2\cdot\left(\frac{1}{2}\right)^2\cdot7\right\rbrace ~=~ 3\mod4\;.
\label{u1anomaly}
\end{split}
\end{align}
\end{subequations}
Here, we use square brackets to highlight the contributions arising from the
$\rep{10}$ and $\crep{5}$--plets, and GUT normalization for the U(1) charges.
There is no contribution from the first and second family of the $\rep{10}$ as
well as from the third family of the $\crep{5}$--plets since their charge
coincides with the superspace charge,  i.e.\ the respective fermions are
uncharged. Note that it is of fundamental importance that the Higgs fields are
in different representations, otherwise the $\Z8^R$ subgroup could \emph{never}
be anomaly universal in this setup (cf.\ the discussion around
\Eqref{eq:higgs_condition}). From the form of the anomaly and equation
\eqref{an_combined} we can immediately conclude that also the $\mathsf{v}^2$,
i.e.\ the unbroken $\Z4^R$ subgroup appears anomalous with 
\begin{equation}
A_{G-G-\Z{4(\mathsf{v}^2)}^R}~=~1\mod2\;.
\end{equation}
Indeed, this anomaly is consistent with the findings of \cite{Lee:2011dya} as it
should be, and the anomalies can be canceled by the Green--Schwarz mechanism.

Let us finally briefly comment on the $\Z4^R$ phenomenology \cite{Lee:2010gv,Lee:2011dya}. The $\Z4^R$ forbids
the $\mu$ term in the MSSM but appears to be broken by non--perturbative
effects. Since the order parameter of $R$ symmetry breaking is the gravitino
mass, a realistic effective $\mu$ term appears. Further, $\Z4^R$ contains $R$ or
matter parity, such that dimension four proton decay operators are forbidden
and dimension five operators are sufficiently suppressed.

As it is known that Abelian discrete $R$ symmetries
\cite{Kappl:2010yu,Bizet:2013gf} and non--Abelian discrete  symmetries
\cite{Kobayashi:2006wq} can originate from orbifold compactifications it is
tempting to speculate that non--Abelian discrete $R$ symmetries may arise in
non--Abelian orbifold compactifications, which have been studied recently in
\cite{Konopka:2012gy,Fischer:2013qza}.

\subsection{Comments on \texorpdfstring{$\boldsymbol{R}$}{R} symmetries and the structure of soft terms}

As is well known, the soft supersymmetry breaking terms are generated by
appropriate effective operators involving a supersymmetry breaking spurion $X$.
Specifically, for the scalar squared masses, the so--called $A$ terms
and the gaugino masses, these operators read schematically (cf.\ e.g.\
\cite{Martin:1997ns})
\begin{subequations}
\begin{eqnarray}
 \int\!\D^4\theta\,\frac{X^\dagger X}{\Lambda^2}\,Q^\dagger Q
 & \xrightarrow{X\to F_X\,\theta^2} & \widetilde{m}^2\,|q|^2
 \;,\label{eq:mtildeSquared}\\
 \int\!\D^2\theta\,\frac{X}{\Lambda}\,y\,Q^3
 & \xrightarrow{X\to F_X\,\theta^2} & A\,y\,q^3\;,
 \label{eq:Aterm}\\
 \int\!\D^2\theta\,\frac{X}{\Lambda}\,W_\alpha W^\alpha
 & \xrightarrow{X\to F_X\,\theta^2} & M_\lambda\,\lambda\lambda\;.
 \label{eq:GauginoMasses}
\end{eqnarray}
\end{subequations}
Here $\Lambda$ is the cut--off scale, $Q$ denotes a generic matter field and
$W_\alpha$ is the multiplet containing the gaugino $\lambda$. If the matter fields
furnish non--trivial representations under a non--Abelian (discrete) symmetry,
one obtains from \eqref{eq:mtildeSquared} soft terms that are, at leading order,
diagonal and get corrected by the
flavor symmetry breaking terms. This leads to a structure that is somewhat similar to the one of
`minimal flavor violation' \cite{Chivukula:1987py,Buras:2000dm} and can help to
ameliorate or solve the supersymmetric flavor problems. 

Let us now entertain the possibility that $X$ has non--zero $R$ charge
under an appropriate, i.e.\ discrete or approximate, $R$ symmetry. In fact, in the simplest scheme of supersymmetry
breaking, such as the Polonyi model and the scenarios of meta--stable
supersymmetry breaking \cite{Intriligator:2006dd}, this situation is realized.
Then the operator \eqref{eq:mtildeSquared} is still allowed while  the $A$ terms
\eqref{eq:Aterm} and gaugino masses \eqref{eq:GauginoMasses} are forbidden.
Since the latter is phenomenologically excluded, one may introduce a second
spurion $X'$ with zero $R$ charge. For $|F_{X}|\gg |F_{X'}|$ one then obtains
heavy scalars and suppressed $A$ terms and gaugino masses. This pattern is also
obtained from KKLT--type moduli stabilization \cite{Kachru:2003aw} with uplift
by a matter field \cite{Lebedev:2006qq}. Here we see that this pattern can be
enforced in a bottom--up approach by imposing $R$ symmetries (but we have no
explanation for the hierarchy $|F_{X}|\gg |F_{X'}|$). This discussion shows that
$R$ symmetries can be instrumental for engineering a certain pattern of soft
terms.

Assume now that there is a non--Abelian discrete non--$R$ symmetry $H$. If
$X$ is to furnish a higher--dimensional representation under $H$, there might be
$H$--invariant contractions between $X$ and the ingredients of the Yukawa
couplings. In this case, provided the $F$--term VEVs of $X$ and the flavon VEVs
are not `aligned', this will generically give rise to very
dangerous flavor--violating operators via \eqref{eq:Aterm}.
On the other hand, if the non--Abelian
symmetry is also an $R$ symmetry, these operators can be forbidden by assigning
a non--zero $R$ charge to the $X$ field. One could then entertain the possibility
that flavor and supersymmetry breaking is due to a single `hidden sector'.
Explicit model building in this direction is, however, beyond the scope of the
present study.

\section{Summary}
\label{sec:Summary}

In this paper we have discussed non--Abelian discrete $R$ symmetries $D$. For
phenomenological reasons we restricted ourselves to settings with
$\mathcal{N}=1$ supersymmetry in which the superspace coordinate $\theta$
furnishes a non--trivial one--dimensional representation of $D$. We have
explored anomalies for such kinds of symmetries. In the course of 
this, we also have shown that perfect groups are always anomaly--free, which 
is of importance especially for the non--$R$ case.
It is instructive to compare GS anomaly cancellation for different kinds of
symmetries. In the case of an Abelian (continuous or discrete) symmetry, one can
always cancel anomalies by the GS mechanism. In the case of a non--Abelian
continuous (gauged) symmetries, an anomaly simply signals an inconsistency.
Finally, for discrete non--Abelian symmetries, there is the possibility of
multiple GS cancellation within one symmetry group. Here one can have
different group operations associated with the shift of different (linear
combinations of) axions. We have worked out the anomaly coefficients
(\Eqref{eq:AnomalyCoefficients}), and discussed GS anomaly cancellation in
detail. 

To illustrate our results, we discussed a toy model in which the MSSM gets
amended by the discrete non--Abelian $R$ symmetry $\Z3\SemiDirect\Z8^R$. The
model combines a flavor symmetry, which dictates certain relations between the
Yukawa couplings, with an $R$ symmetry that suppresses the $\mu$ term and
dangerous proton decay operators. Moreover, due to the fact that it is an $R$
symmetry, representations for so--called driving fields are automatically present in the spectrum, hence the
question of `VEV alignment' can be addressed without enlarging the symmetry
group. Although the toy model is certainly not fully realistic, it illustrates
the novel possibilities that arise once one promotes ordinary non--Abelian flavor
symmetries to $R$ symmetries: one can address the question of flavor and
simultaneously solve the proton decay and
$\mu$ problems with a single symmetry.

\section*{Acknowledgments}

We would like to thank Maximilian Fallbacher and Patrick Vaudrevange for
useful discussions. M.-C.C.\ would like to thank TU M\"unchen, where part of the
work  was done, for hospitality. M.R.\ would like to thank the  UC Irvine, where
part of this work was done, for  hospitality. This work was partially supported
by the DFG cluster  of excellence ``Origin and Structure of the Universe'' and
the Graduiertenkolleg ``Particle Physics at the Energy Frontier of New
Phenomena'' by Deutsche Forschungsgemeinschaft (DFG). The work of M.-C.C.\ was
supported, in part, by the U.S.\ National Science  Foundation under Grant No.\
PHY-0970173.  M.-C.C.\ and  M.R.\  would like to thank CETUP* for  hospitality
and support. This research was done in the context of the ERC  Advanced Grant
project ``FLAVOUR''~(267104).

\appendix

\section{The group \texorpdfstring{$\boldsymbol{\Z3\SemiDirect\Z8}$}{Z3xZ8}}
\label{appsec:group}

Let us briefly describe the relevant features of the group
$\Z3\SemiDirect\Z8$. A presentation of the group has already been given in \eqref{eq:presentation}. The character table is given in \ref{tab:char}. 
\begin{table}[t!]
\centering
\begin{tabular}{c|rrrrrrrrrrrr|}
                      &  $\mathbbm{1}$ & $\mathsf{v}$ & $\mathsf{v}^2$ & $\mathsf{v}^4$ & $\mathsf{u}$ & $\mathsf{v}^3$ & $\mathsf{v}^5$ & $\mathsf{v}^6$ & $\mathsf{uv}^2$ & $\mathsf{uv}^4$ & $\mathsf{v}^7$ & $\mathsf{uv}^6$ \\
                      &  1 &  3 &  1 &  1 &  2 &  3 &  3 &  1 &   2 &  2 &  3 &   2 \\
$\Z3\SemiDirect\Z8$       & 1a & 8a & 4a & 2a & 3a & 8b & 8c & 4b & 12a & 6a & 8d & 12b \\
\hline
 $\rep{1}_1$ & $1$ &  $1$ &  $1$ &  $1$ &  $1$ &  $1$ &  $1$ &  $1$ &  $1$ &  $1$ &  $1$ &  $1$ \\
 $\rep{1}_2$ & $1$ & $-1$ &  $1$ &  $1$ &  $1$ & $-1$ & $-1$ &  $1$ &  $1$ &  $1$ & $-1$ &  $1$ \\
 $\rep{1}_3$ & $1$ & $-\I$ & $-1$ &  $1$ &  $1$ &  $\I$ & $-\I$ & $-1$ & $-1$ &  $1$ &  $\I$ & $-1$ \\
 $\rep{1}_4$ & $1$ &  $\I$ & $-1$ &  $1$ &  $1$ & $-\I$ &  $\I$ & $-1$ & $-1$ &  $1$ & $-\I$ & $-1$ \\
 $\rep{1}_5$ & $1$ &   $-\tau$ &  $\I$ & $-1$ & $1$ &  $\tau^*$ &    $\tau$ & $-\I$ &  $\I$ & $-1$ & $-\tau^*$ & $-\I$ \\
 $\rep{1}_6$ & $1$ &  $\tau^*$ & $-\I$ & $-1$ & $1$ &   $-\tau$ & $-\tau^*$ &  $\I$ & $-\I$ & $-1$ &    $\tau$ &  $\I$ \\
 $\rep{1}_7$ & $1$ & $-\tau^*$ & $-\I$ & $-1$ & $1$ &    $\tau$ &  $\tau^*$ &  $\I$ & $-\I$ & $-1$ &   $-\tau$ &  $\I$ \\
 $\rep{1}_8$ & $1$ &    $\tau$ &  $\I$ & $-1$ & $1$ & $-\tau^*$ &   $-\tau$ & $-\I$ &  $\I$ & $-1$ &  $\tau^*$ & $-\I$ \\
 $\rep{2}_1$ & $2$ &  $0$ & $-2$ & $2$ & $-1$ & $0$ & $0$ & $-2$ &  $1$ & $-1$ & $0$ &  $1$ \\
 $\rep{2}_2$ & $2$ &  $0$ &  $2$ & $2$ & $-1$ & $0$ & $0$ &  $2$ & $-1$ & $-1$ & $0$ & $-1$ \\
 $\rep{2}_3$ & $2$ &  $0$ & $-2\I$ & $-2$ & $-1$ & $0$ & $0$ &  $2\I$ &  $\I$ & $1$ & $0$ & $-\I$ \\
 $\rep{2}_4$ & $2$ &  $0$ &  $2\I$ & $-2$ & $-1$ & $0$ & $0$ & $-2\I$ & $-\I$ & $1$ & $0$ &  $\I$ \\
\hline
\end{tabular}
\caption{Character table of $\Z3\SemiDirect\Z8$. We define $\tau:=\e^{\nicefrac{2\pi\I}{8}}$. 
The conjugacy classes (c.c.) are labeled by the order of their elements and a letter. 
The first line gives a representative of the c.c. in the presentation specified in the text. 
The second line gives the cardinality of the corresponding c.c. }
\label{tab:char}
\end{table}
The product rules for the irreducible representations are stated in tables~\ref{tab:products1} and
\ref{tab:products2}. 
\begin{table}[t]
\centering
\begin{tabular}{c|ccccccccccc|}
 $\otimes$ &$\rep{1}_2$&$\rep{1}_3$&$\rep{1}_4$&$\rep{1}_5$&$\rep{1}_6$&$\rep{1}_7$&$\rep{1}_8$&$\rep{2}_1$&$\rep{2}_2$&$\rep{2}_3$&$\rep{2}_4$ \\
\hline
 $\rep{1}_2$ &$\rep{1}_1$&$\rep{1}_4$&$\rep{1}_3$&$\rep{1}_8$&$\rep{1}_7$&$\rep{1}_6$&$\rep{1}_5$ &$\rep{2}_1$&$\rep{2}_2$&$\rep{2}_3$&$\rep{2}_4$ \\
 $\rep{1}_3$ &$\rep{1}_4$&$\rep{1}_2$&$\rep{1}_1$&$\rep{1}_7$&$\rep{1}_5$&$\rep{1}_8$&$\rep{1}_6$ &$\rep{2}_2$&$\rep{2}_1$&$\rep{2}_4$&$\rep{2}_3$ \\
 $\rep{1}_4$ &$\rep{1}_3$&$\rep{1}_1$&$\rep{1}_2$&$\rep{1}_6$&$\rep{1}_8$&$\rep{1}_5$&$\rep{1}_7$ &$\rep{2}_2$&$\rep{2}_1$&$\rep{2}_4$&$\rep{2}_3$ \\
 $\rep{1}_5$ &$\rep{1}_8$&$\rep{1}_7$&$\rep{1}_6$&$\rep{1}_4$&$\rep{1}_2$&$\rep{1}_1$&$\rep{1}_3$ &$\rep{2}_3$&$\rep{2}_4$&$\rep{2}_2$&$\rep{2}_1$ \\
 $\rep{1}_6$ &$\rep{1}_7$&$\rep{1}_5$&$\rep{1}_8$&$\rep{1}_2$&$\rep{1}_3$&$\rep{1}_4$&$\rep{1}_1$ &$\rep{2}_4$&$\rep{2}_3$&$\rep{2}_1$&$\rep{2}_2$ \\
 $\rep{1}_7$ &$\rep{1}_6$&$\rep{1}_8$&$\rep{1}_5$&$\rep{1}_1$&$\rep{1}_4$&$\rep{1}_3$&$\rep{1}_2$ &$\rep{2}_4$&$\rep{2}_3$&$\rep{2}_1$&$\rep{2}_2$ \\
 $\rep{1}_8$ &$\rep{1}_5$&$\rep{1}_6$&$\rep{1}_7$&$\rep{1}_3$&$\rep{1}_1$&$\rep{1}_2$&$\rep{1}_4$ &$\rep{2}_3$&$\rep{2}_4$&$\rep{2}_2$&$\rep{2}_1$ \\

 $\rep{2}_1$ &$\rep{2}_1$&$\rep{2}_2$&$\rep{2}_2$&$\rep{2}_3$&$\rep{2}_4$&$\rep{2}_4$&$\rep{2}_3$ &$ $&$ $&$ $&$ $ \\
 $\rep{2}_2$ &$\rep{2}_2$&$\rep{2}_1$&$\rep{2}_1$&$\rep{2}_4$&$\rep{2}_3$&$\rep{2}_3$&$\rep{2}_4$ &$ $&$ $&$ $&$ $ \\
 $\rep{2}_3$ &$\rep{2}_3$&$\rep{2}_4$&$\rep{2}_4$&$\rep{2}_2$&$\rep{2}_1$&$\rep{2}_1$&$\rep{2}_2$ &$ $&$ $&$ $&$ $ \\
 $\rep{2}_4$ &$\rep{2}_4$&$\rep{2}_3$&$\rep{2}_3$&$\rep{2}_1$&$\rep{2}_2$&$\rep{2}_2$&$\rep{2}_1$ &$ $&$ $&$ $&$ $ \\
\hline
\end{tabular}
\caption{Decomposition of the tensor products of irreducible representations of one--dimensional representations
and doublets with one--dimensional representations.}
\label{tab:products1}
\end{table}
\begin{table}[t]
\centering
\begin{tabular}{c|cccc|}
 $\otimes$ &$\rep{2}_1$&$\rep{2}_2$&$\rep{2}_3$&$\rep{2}_4$ \\
\hline
 $\rep{2}_1$ & $\rep1_1\oplus\rep1_2\oplus\rep{2}_2$ & $\rep1_3\oplus\rep1_4\oplus\rep{2}_1$ & $\rep1_5\oplus\rep1_8\oplus\rep{2}_4$ & $\rep1_6\oplus\rep1_7\oplus\rep{2}_3$ \\
 $\rep{2}_2$ & $\rep1_3\oplus\rep1_4\oplus\rep{2}_1$ & $\rep1_1\oplus\rep1_2\oplus\rep{2}_2$ & $\rep1_6\oplus\rep1_7\oplus\rep{2}_3$ & $\rep1_5\oplus\rep1_8\oplus\rep{2}_4$ \\
 $\rep{2}_3$ & $\rep1_5\oplus\rep1_8\oplus\rep{2}_4$ & $\rep1_6\oplus\rep1_7\oplus\rep{2}_3$ & $\rep1_3\oplus\rep1_4\oplus\rep{2}_1$ & $\rep1_1\oplus\rep1_2\oplus\rep{2}_2$ \\
 $\rep{2}_4$ & $\rep1_6\oplus\rep1_7\oplus\rep{2}_3$ & $\rep1_5\oplus\rep1_8\oplus\rep{2}_4$ & $\rep1_1\oplus\rep1_2\oplus\rep{2}_2$ & $\rep1_3\oplus\rep1_4\oplus\rep{2}_1$ \\
\hline
\end{tabular}
\caption{Decomposition of the tensor products of two doublet representations.}
\label{tab:products2}
\end{table}
For the doublet representations
$\rep{2}_j$, a possible form of the $\Z3$ and $\Z8$ generators $\mathsf{u}$ and
$\mathsf{v}$ is given by
\begin{subequations}
\begin{align}
 \widetilde{U}_j& ~=~ \widetilde{U}~=~\frac{1}{2}\begin{pmatrix} -1 & \I\,\sqrt{3} \\ \I\,\sqrt{3} & -1 \end{pmatrix}\;,
\label{eq:matrix_rep_Z8_A}\\
\widetilde{V}_1 & ~=~  \begin{pmatrix} \I & 0 \\ 0 & -\I \end{pmatrix}\;,\qquad
\widetilde{V}_2~=~\begin{pmatrix} 1 & 0 \\ 0 & -1 \end{pmatrix}\;,\nonumber\\
\widetilde{V}_3& ~=~\begin{pmatrix} \tau^* & 0 \\ 0 & -\tau^* \end{pmatrix}
 \quad\text{and}\quad
\widetilde{V}_4~=~\begin{pmatrix} \tau & 0 \\ 0 & -\tau \end{pmatrix}\;.
\label{eq:matrix_rep_Z8_B}
\end{align}\label{eq:matrix_rep_Z8}
\end{subequations}
Here we have used $\tau:=\e^{\nicefrac{2\pi\I}{8}}$ to denote the eight root of unity.
We also state the explicit form of all the tensor products which one may need for the construction of the mass matrices of possible models. 
Let $(a_1,a_2)^T$ and $(b_1,b_2)^T$ each transform as a doublet
and $c$ be a one--dimensional representation. Then
\begin{subequations}\label{eq:tensorsZ8B}
\begin{align}
(a_{\rep{2}_4}\otimes c_{\rep{1}_5}) & ~=~  \binom{a_2\,c}{a_1\,c}_{\rep{2}_1}\;,\\
(a_{\rep{2}_4}\otimes c_{\rep{1}_8}) & ~=~  \binom{a_1\,c}{a_2\,c}_{\rep{2}_1}\;,\label{tensor1b}\\
(a_{\rep{2}_2}\otimes b_{\rep{2}_1}) & ~=~ \left(a_1\,b_2-a_2\,b_1\right)_{\rep{1}_3}
\oplus\left(a_1\,b_1-\,a_2\,b_2\right)_{\rep{1}_4}\oplus\binom{a_1\,b_1+a_2\,b_2}{-(a_1\,b_2+a_2\,b_1)}_{\rep{2}_1}\;,\label{tensor4b}\\
(a_{\rep{2}_2}\otimes b_{\rep{2}_4}) & ~=~ \left(a_1\,b_2-a_2\,b_1\right)_{\rep{1}_5}
\oplus\left(a_1\,b_1-\,a_2\,b_2\right)_{\rep{1}_8}\oplus\binom{a_1\,b_1+a_2\,b_2}{-(a_1\,b_2+a_2\,b_1)}_{\rep{2}_4}\;,\\
(a_{\rep{2}_4}\otimes b_{\rep{2}_4}) & ~=~ \left(a_1\,b_2-a_2\,b_1\right)_{\rep{1}_3}
\oplus\left(a_1\,b_1-\,a_2\,b_2\right)_{\rep{1}_4}\oplus\binom{a_1\,b_1+a_2\,b_2}{-(a_1\,b_2+a_2\,b_1)}_{\rep{2}_1}\;,\label{tensor5b}\\
(a_{\rep{2}_2}\otimes b_{\rep{2}_2}) & ~=~ \left(a_1\,b_1-a_2\,b_2\right)_{\rep{1}_1}
\oplus(a_1\,b_2-a_2\,b_1)_{\rep{1}_2}\oplus\binom{a_1\,b_1+a_2\,b_2}{-(a_1\,b_2+a_2\,b_1)}_{\rep{2}_2}\;.\label{tensor6b}
\end{align}
\end{subequations}

\bibliography{Orbifold}

\providecommand{\bysame}{\leavevmode\hbox to3em{\hrulefill}\thinspace}
\frenchspacing
\newcommand{\origttfamily}{}
\let\origttfamily=\ttfamily
\renewcommand{\ttfamily}{\origttfamily \hyphenchar\font=`\-}

\begin{thebibliography}{10}

\bibitem{Krauss:1988zc}
L.~M. Krauss and F.~Wilczek, Phys. Rev. Lett. \textbf{62} (1989), 1221.

\bibitem{Ibanez:1991hv}
L.~E. Ib{\'a}{\~n}ez and G.~G. Ross, Phys. Lett. \textbf{B260} (1991), 291.

\bibitem{Ibanez:1991pr}
L.~E. Ib{\'a}{\~n}ez and G.~G. Ross, Nucl. Phys. \textbf{B368} (1992), 3.

\bibitem{Banks:1991xj}
T.~Banks and M.~Dine, Phys. Rev. \textbf{D45} (1992), 1424,
  \texttt{hep-th/9109045}.

\bibitem{Fujikawa:1979ay}
K.~Fujikawa, Phys. Rev. Lett. \textbf{42} (1979), 1195.

\bibitem{Fujikawa:1980eg}
K.~Fujikawa, Phys. Rev. \textbf{D21} (1980), 2848.

\bibitem{Araki:2006mw}
T.~Araki, Prog. Theor. Phys. \textbf{117} (2007), 1119,
  \texttt{hep-ph/0612306}.

\bibitem{Araki:2008ek}
T.~Araki et~al., Nucl. Phys. \textbf{B805} (2008), 124,
  \texttt{arXiv:0805.0207} [hep-th].

\bibitem{Chen:2012jg}
M.-C. Chen, M.~Ratz, C.~Staudt, and P.~K. Vaudrevange, Nucl.Phys. \textbf{B866}
  (2012), 157, \texttt{arXiv:1206.5375} [hep-ph].

\bibitem{Lee:2011dya}
H.~M. Lee, S.~Raby, M.~Ratz, G.~G. Ross, R.~Schieren, K.~Schmidt-Hoberg, and
  P.~K. Vaudrevange, Nucl.Phys. \textbf{B850} (2011), 1,
  \texttt{arXiv:1102.3595} [hep-ph].

\bibitem{Ludeling:2012cu}
C.~L{\"u}deling, F.~Ruehle, and C.~Wieck, Phys.Rev. \textbf{D85} (2012),
  106010, \texttt{arXiv:1203.5789} [hep-th].

\bibitem{Chen:2012pi}
M.-C. Chen, M.~Fallbacher, and M.~Ratz, Mod.Phys.Lett. \textbf{A27} (2012),
  1230044, \texttt{arXiv:1211.6247} [hep-ph].

\bibitem{Lee:2010gv}
H.~M. Lee, S.~Raby, M.~Ratz, G.~G. Ross, R.~Schieren, K.~Schmidt-Hoberg, and
  P.~K. Vaudrevange, Phys.Lett. \textbf{B694} (2011), 491,
  \texttt{arXiv:1009.0905} [hep-ph].

\bibitem{Ramond:2010zz}
P.~Ramond, \emph{Group theory: A physicist's survey}, 2010.

\bibitem{Babu:2002tx}
K.~S. Babu, I.~Gogoladze, and K.~Wang, Nucl. Phys. \textbf{B660} (2003), 322,
  \texttt{hep-ph/0212245}.

\bibitem{GAP4}
The GAP~Group, \emph{{GAP -- Groups, Algorithms, and Programming, Version
  4.5.5}}, 2012.

\bibitem{Chen:2004rr}
S.-L. Chen, M.~Frigerio, and E.~Ma, Phys.Rev. \textbf{D70} (2004), 073008,
  \texttt{arXiv:hep-ph/0404084} [hep-ph].

\bibitem{Morisi:2005fy}
S.~Morisi and M.~Picariello, Int.J.Theor.Phys. \textbf{45} (2006), 1267,
  \texttt{arXiv:hep-ph/0505113} [hep-ph].

\bibitem{Dermisek:2005ij}
R.~Dermisek and S.~Raby, Phys.Lett. \textbf{B622} (2005), 327,
  \texttt{arXiv:hep-ph/0507045} [hep-ph].

\bibitem{Caravaglios:2005gf}
F.~Caravaglios and S.~Morisi, \texttt{arXiv:hep-ph/0510321} [hep-ph].

\bibitem{Teshima:2005bk}
T.~Teshima, Phys.Rev. \textbf{D73} (2006), 045019,
  \texttt{arXiv:hep-ph/0509094} [hep-ph].

\bibitem{Haba:2005ds}
N.~Haba and K.~Yoshioka, Nucl.Phys. \textbf{B739} (2006), 254,
  \texttt{arXiv:hep-ph/0511108} [hep-ph].

\bibitem{Picariello:2006sp}
M.~Picariello, Int.J.Mod.Phys. \textbf{A23} (2008), 4435,
  \texttt{arXiv:hep-ph/0611189} [hep-ph].

\bibitem{Mohapatra:2006un}
R.~Mohapatra, S.~Nasri, and H.-B. Yu, Phys.Lett. \textbf{B636} (2006), 114,
  \texttt{arXiv:hep-ph/0603020} [hep-ph].

\bibitem{Mohapatra:2006pu}
R.~Mohapatra, S.~Nasri, and H.-B. Yu, Phys.Lett. \textbf{B639} (2006), 318,
  \texttt{arXiv:hep-ph/0605020} [hep-ph].

\bibitem{Feruglio:2007hi}
F.~Feruglio and Y.~Lin, Nucl.Phys. \textbf{B800} (2008), 77,
  \texttt{arXiv:0712.1528} [hep-ph].

\bibitem{Antusch:2005gp}
S.~Antusch, J.~Kersten, M.~Lindner, M.~Ratz, and M.~A. Schmidt, JHEP
  \textbf{03} (2005), 024, \texttt{hep-ph/0501272}.

\bibitem{Altarelli:2010gt}
G.~Altarelli and F.~Feruglio, Rev.Mod.Phys. \textbf{82} (2010), 2701,
  \texttt{arXiv:1002.0211} [hep-ph].

\bibitem{Ishimori:2010au}
H.~Ishimori, T.~Kobayashi, H.~Ohki, Y.~Shimizu, H.~Okada, et~al.,
  Prog.Theor.Phys.Suppl. \textbf{183} (2010), 1, \texttt{arXiv:1003.3552}
  [hep-th].

\bibitem{King:2013eh}
S.~F. King and C.~Luhn, \texttt{arXiv:1301.1340} [hep-ph].

\bibitem{Kappl:2010yu}
R.~Kappl, B.~Petersen, S.~Raby, M.~Ratz, R.~Schieren, and P.~K. Vaudrevange,
  Nucl.Phys. \textbf{B847} (2011), 325, \texttt{arXiv:1012.4574} [hep-th].

\bibitem{Bizet:2013gf}
N.~G.~C. Bizet, T.~Kobayashi, D.~K.~M. Pena, S.~L. Parameswaran, M.~Schmitz,
  et~al., \texttt{arXiv:1301.2322} [hep-th].

\bibitem{Kobayashi:2006wq}
T.~Kobayashi, H.~P. Nilles, F.~Pl{\"o}ger, S.~Raby, and M.~Ratz, Nucl. Phys.
  \textbf{B768} (2007), 135, \texttt{hep-ph/0611020}.

\bibitem{Konopka:2012gy}
S.~J. Konopka, \texttt{arXiv:1210.5040} [hep-th].

\bibitem{Fischer:2013qza}
M.~Fischer, S.~Ramos-S{\'a}nchez, and P.~K.~S. Vaudrevange,
  \texttt{arXiv:1304.7742} [hep-th].

\bibitem{Martin:1997ns}
S.~P. Martin, \texttt{hep-ph/9709356}.

\bibitem{Chivukula:1987py}
R.~S. Chivukula and H.~Georgi, Phys. Lett. \textbf{B188} (1987), 99.

\bibitem{Buras:2000dm}
A.~J. Buras, P.~Gambino, M.~Gorbahn, S.~J{\"a}ger, and L.~Silvestrini, Phys.
  Lett. \textbf{B500} (2001), 161, \texttt{hep-ph/0007085}.

\bibitem{Intriligator:2006dd}
K.~Intriligator, N.~Seiberg, and D.~Shih, JHEP \textbf{04} (2006), 021,
  \texttt{hep-th/0602239}.

\bibitem{Kachru:2003aw}
S.~Kachru, R.~Kallosh, A.~Linde, and S.~P. Trivedi, Phys. Rev. \textbf{D68}
  (2003), 046005, \texttt{hep-th/0301240}.

\bibitem{Lebedev:2006qq}
O.~Lebedev, H.~P. Nilles, and M.~Ratz, Phys. Lett. \textbf{B636} (2006), 126,
  \texttt{hep-th/0603047}.

\end{thebibliography}
\addcontentsline{toc}{section}{Bibliography}
\bibliographystyle{NewArXiv} 
\end{document}